\documentclass[conference]{IEEEtran}
\IEEEoverridecommandlockouts

\usepackage{cite}
\usepackage{amsmath,amssymb,amsfonts}
\usepackage{graphicx}
\usepackage{textcomp}
\usepackage{subcaption}
\usepackage{lastpage}
\usepackage{placeins}
\usepackage{makecell}
\usepackage{enumitem}
\usepackage{booktabs}
\usepackage{multirow}
\usepackage{array}
\usepackage{algorithm}
\usepackage{xcolor}
\usepackage{hyperref}
\usepackage{algpseudocode}
\def\BibTeX{{\rm B\kern-.05em{\sc i\kern-.025em b}\kern-.08em
    T\kern-.1667em\lower.7ex\hbox{E}\kern-.125emX}}
\begin{document}

\title{Low-Data Classification of Historical Music Manuscripts: A Few-Shot Learning Approach\\  

}

\author{
\IEEEauthorblockN{
Elona Shatri, 
Daniel Raymond, 
Gy\"orgy Fazekas
}
\IEEEauthorblockA{\textit{Centre for Digital Music} \\
\textit{Queen Mary University of London}\\
London, UK \\
\{e.shatri, d.k.raymond, george.fazekas\}@qmul.ac.uk}
}

\maketitle
\begin{abstract} 

In this paper, we explore the intersection of technology and cultural preservation by developing a self-supervised learning framework for the classification of musical symbols in historical manuscripts. Optical Music Recognition (OMR) plays a vital role in digitising and preserving musical heritage, but historical documents often lack the labelled data required by traditional methods. We overcome this challenge by training a neural-based feature extractor on unlabelled data, enabling effective classification with minimal samples. Key contributions include optimising crop preprocessing for a self-supervised Convolutional Neural Network and evaluating classification methods, including SVM, multilayer perceptrons, and prototypical networks. Our experiments yield an accuracy of 87.66\%, showcasing the potential of AI-driven methods to ensure the survival of historical music for future generations through advanced digital archiving techniques.
\end{abstract}

\section{Introduction}\label{sec:intro}

The digitisation of historical sheet music is crucial for preserving musical heritage. Optical Music Recognition (OMR) techniques convert these documents into machine-readable formats like MusicXML and MIDI, allowing integration into modern platforms. However, recognising symbols in these manuscripts presents challenges due to document degradation and variability \cite{shatri_optical_2020}.

Existing deep learning approaches to OMR typically rely on large, annotated datasets, but the scarcity of labelled data for historical documents presents a major obstacle \cite{castellanos_preliminary_2023}. Without sufficient labels, traditional OMR systems struggle to generalise across diverse notational styles found in historical archives. This paper addresses these limitations by proposing a self-supervised learning framework that facilitates robust music symbol classification with minimal labelled data, thus supporting the broader goal of preserving and digitising musical heritage.

Building on the foundations of previous research by \cite{alfaro-contreras_few-shot_2023}, we hypothesise that a self-supervised Convolutional Neural Network (CNN), paired with optimised classification algorithms, can effectively classify music symbols using only a few annotated examples. A key focus of this work is optimising the preprocessing of musical "crops"—small sections of manuscripts—where we apply transformations that enhance the model’s robustness to the varied forms of degradation typically found in historical documents. The extracted features are subsequently classified using methods like Support Vector Machines (SVM), multilayer perceptrons (MLP), and prototypical networks, ensuring accuracy despite limited sample sizes.

Through experiments on historical music manuscripts, we demonstrate the advantages of this self-supervised pipeline over traditional supervised approaches.

\begin{figure*}[t]
    \centering
    \includegraphics[width=0.85\textwidth]{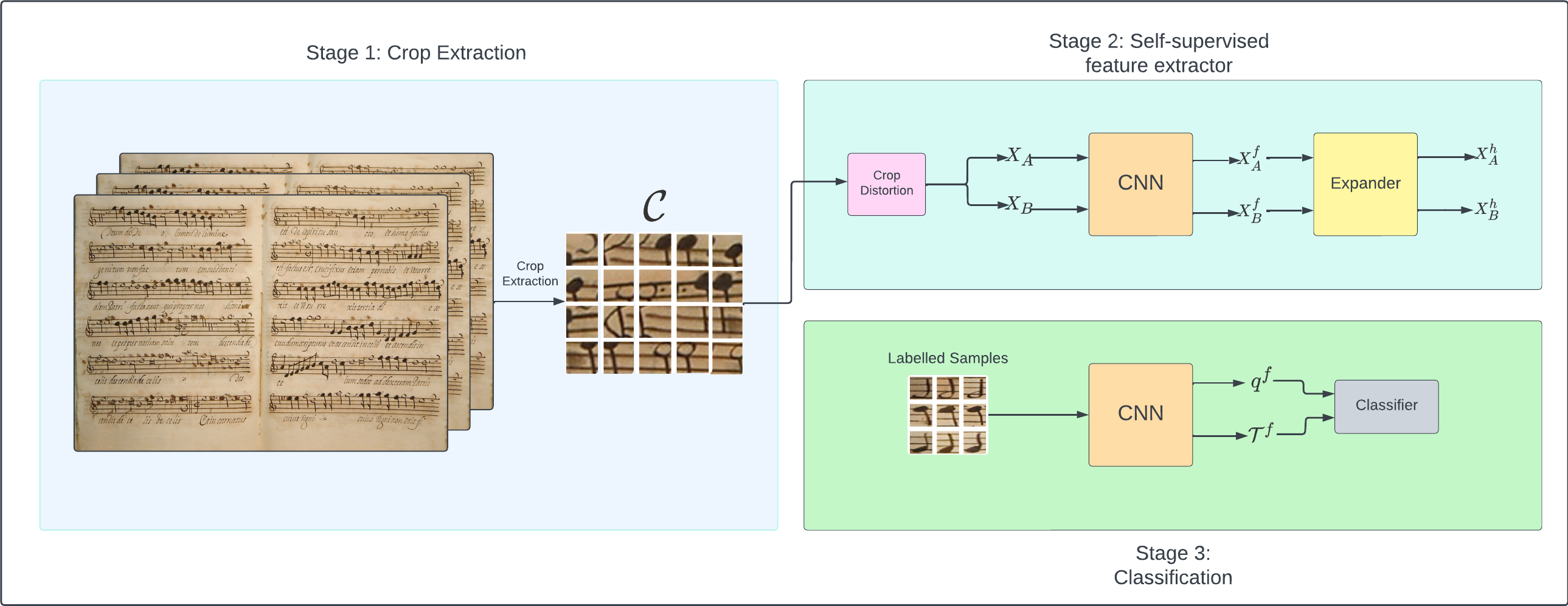}
    \caption{A depiction of the musical symbol classification pipeline based on the pipeline by \cite{alfaro-contreras_few-shot_2023}.}
    \label{fig:pipeline}
\end{figure*}

\section{Background}

As mentioned in Section \ref{sec:intro}, the digitisation of cultural artefacts, such as historical sheet music, is an essential component of modern efforts to preserve and share cultural heritage. However, this process presents significant challenges, particularly in fields like OMR, where the diversity and degradation of historical documents require more advanced approaches than those used for modern printed music. Deep learning models have proven effective in image classification tasks, but they typically rely on large volumes of annotated data—data that is often scarce for historical manuscripts \cite{wang_generalizing_2021}.

Few-shot learning addresses this limitation by enabling models to generalise from only a few training examples. This is particularly valuable for tasks like OMR in historical contexts, where obtaining extensive labelled data is impractical. The few-shot learning paradigm, often utilising an N-way-K-shot approach, empowers models to recognise and classify symbols based on a minimal number of labelled samples \cite{wang_generalizing_2021}. Current OMR systems for modern, printed scores achieve high accuracy through deep learning techniques such as transformer \cite{rios2024} and object detection \cite{shatri2021worms,shatri2024kdir} largely because of the uniformity and well-annotated nature of the data. However, when applied to handwritten or historical music manuscripts, these models struggle due to variability in musical notation, handwriting styles, and document degradation over time \cite{shatri_optical_2020}.

To mitigate these challenges, few-shot learning techniques can be divided into two broad categories: metric-based methods and meta-learning approaches. Metric-based methods, such as Siamese networks, project samples into a feature space where similar items cluster together by minimising the distance between them \cite{song_comprehensive_2022}. Prototypical networks extend this concept by computing a prototype for each class, allowing new samples to be classified based on their proximity to these prototypes \cite{snell_prototypical_2017}. These approaches are well-suited for the task of recognising symbols in historical music manuscripts, where only a few labelled examples may be available.

Meta-learning techniques like Model-Agnostic Meta-Learning (MAML) offer another promising avenue for improving OMR in cultural preservation. These methods enable models to adapt quickly to new tasks with just a few gradient updates, making them ideal for scenarios where the data is highly variable, such as in historical document recognition \cite{hospedales_meta-learning_2022}. Techniques like Reptile optimise across many tasks, improving the model’s generalisation capability, which is crucial for handling the inconsistencies found in historical sheet music \cite{finn_model-agnostic_2017}.

By applying these few-shot and meta-learning techniques to OMR, this study aims to bridge the gap between technological advancements in image classification and the broader goal of preserving musical heritage. The ability to effectively classify symbols with minimal labelled data ensures that even rare and deteriorating manuscripts can be digitised, analysed, and made accessible for future generations, contributing to the ongoing efforts to safeguard cultural history.

\section{Methodology}

This study builds on the work by \cite{alfaro-contreras_few-shot_2023}, adapting a self-supervised Convolutional Neural Network (CNN) to encode relevant features for music symbol classification in historical manuscripts using few-shot learning. Our goal is to provide a framework that not only achieves high classification accuracy but also aids in the digital preservation of musical heritage by overcoming the limitations of scarce labelled data. The key stages of the pipeline are detailed below (see Figure \ref{fig:pipeline}:

\subsection{Crop Extraction}

Historical music sheets often exhibit varying symbol sizes, spacing, and document degradation, making it essential to preprocess the documents carefully before classification. To extract individual musical symbols, we divided the manuscripts into subsections or crops, each ideally containing a single symbol. This was done using a sliding window approach, which processes the manuscript in overlapping segments. Binarisation and entropy-based algorithms filtered out blank or irrelevant sections, improving computational efficiency. We employed the Sauvola method \cite{sauvola_adaptive_1997} for adaptive binarisation to handle the uneven lighting and degradation typically found in historical documents.

\subsection{Self-supervised neural network feature extractor}

Since historical manuscripts often lack extensive labelled data, self-supervised learning enables feature extraction from unlabelled data. We applied the VICReg method \cite{bardes_vicreg_2022}, which trains the CNN by distorting each crop twice and ensuring that these distortions map to the same point in the feature space. This approach is well-suited for handling the variability in symbol appearance caused by document ageing and handwritten inconsistencies. To further enhance network convergence, an expander block was included, and the loss function was structured to balance variance, invariance, and covariance terms, promoting robust feature learning.

Once trained, the CNN produced a high-dimensional feature vector for each symbol, which was subsequently classified using methods such as SVM, MLP, and prototypical networks. These classifiers were chosen for their complementary strengths in handling high-dimensional and few-shot data.

Key improvements were made to both the image transformation and classification stages to enhance robustness and accuracy:

\begin{itemize}
    \item We tested additional transformations, such as salt-and-pepper noise, elastic distortion, and fade, simulating typical degradation in historical manuscripts \cite{shorten_survey_2019}. This ensured consistent feature extraction across varying image qualities.

    \item The pipeline originally used k-nearest neighbors (kNN) for classification, but we experimented with SVM, MLP, and prototypical networks, which better suited the high-dimensional feature space produced by the CNN.
\end{itemize}

\subsection{Crop preprocessing}

To improve the robustness of the feature extraction process, we applied several image transformations, simulating the typical forms of degradation seen in historical music manuscripts. The final set of transformations included random resized crops, greyscale, colour jitter, salt-and-pepper noise, Gaussian blur, and fade. These transformations ensure that the CNN is able to extract consistent features, regardless of the quality of the manuscript. Horizontal flipping was intentionally excluded, as this could introduce ambiguities in symbol orientation, which is critical in music notation.

\subsection{Classification}

We evaluated several classification algorithms, with MLP outperforming others. The MLP was configured with five layers, reducing the input feature dimensions from 1600 to 128 through gradual dimensionality reduction. Regularisation was applied at each stage using batch normalisation and dropout to prevent overfitting, which is particularly important given the high-dimensional feature space generated by the CNN. Additionally, a prototypical classification layer was explored, with additional fully connected layers to improve the embedding’s adaptability, though MLP consistently showed superior performance.

\begin{figure}[ht]
    \centering
    \includegraphics[width=1\columnwidth]{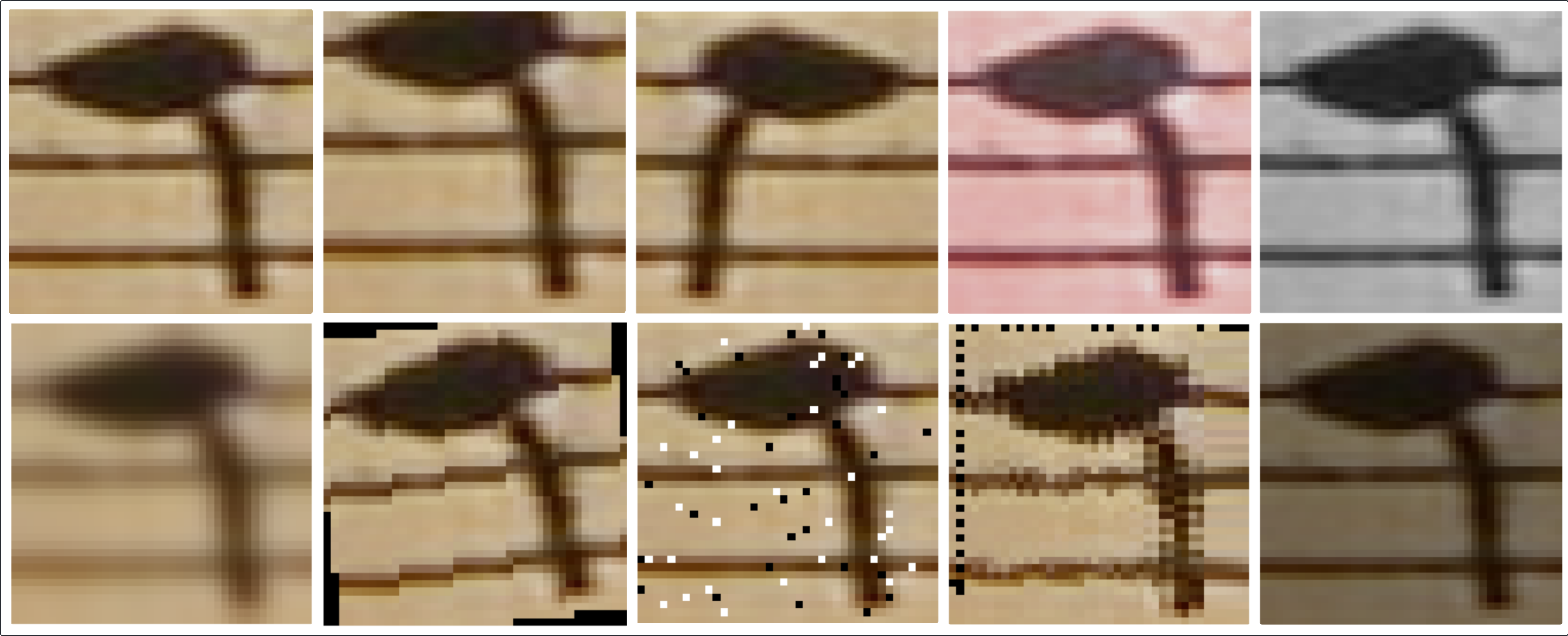}
    \caption{Examples of the transformations utilised prior to training. From left to right are Random resizing, Random horizontal flip, Colour jitter, Random greyscale, Gaussian blur, Random rotation, Salt and pepper noise, Elastic distortion and Fade.}
    \label{fig:augmentation-examples}
\end{figure}

\subsection{Data Augmentation}

To further enhance the classifiers' generalisation capabilities, we employed various data augmentation techniques during training (see Section \ref{sec:expsetup} for details), as shown in Figure \ref{fig:augmentation-examples}. These transformations ensured that the CNN extracted consistent features regardless of document quality.

\section{Experimental Setup}\label{sec:expsetup}

In this section, we outline the dataset used, the pipeline configuration, and the application of data augmentation during the classification stage. The setup was designed to ensure that the model can effectively handle the variability and degradation typical of historical music manuscripts, contributing to their long-term preservation through accurate digital recognition.

We employed the Capitan dataset \cite{calvo-zaragoza_handwritten_2019}, which contains mensural notation extracted from historical music manuscripts. The dataset comprises 98 pages, totalling 17,112 labelled music symbols across 28 distinct classes. This dataset presents a realistic challenge, as the symbols exhibit variability in style and quality due to the age and condition of the manuscripts. Its diversity makes it an ideal testbed for evaluating few-shot learning techniques in the context of cultural preservation.

\subsection{Configuration}
To ensure consistency with prior work and allow for meaningful comparison, we maintained a similar pipeline configuration. The crop extraction process was performed using a sliding window of 64 pixels, with an entropy threshold of 0.8 to filter out blank or uninformative regions. The self-supervised CNN generated a 1600-dimensional feature vector for each crop, using VICReg regularisation parameters set to $\lambda = 10$, $\mu = 10$, and $\phi = 1$ \cite{nunez-alcover_glyph_2019}.

For classification, we used the N-way-K-shot method, which allows the model to generalise from a small number of labeled samples. We experimented with K values of 1, 2, 5, and 10 to simulate varying degrees of data scarcity. The prototypical classifier was tested with different combinations of support ($S$) and query ($Q$) sets: $S=1$, $Q=2$; $S=2$, $Q=3$; and $S=6$, $Q=4$. These configurations were chosen to assess the model's ability to handle different levels of training data availability, which is critical for digitising rare historical documents.

\subsection{Data Augmentation} 

Data augmentation is crucial in improving the model’s robustness to the wide range of image qualities found in historical manuscripts. We applied augmentation at different levels ($A = [0, 1, 2, 5, 10, 20]$), where $A$ represents the number of augmented samples generated per original crop. The transformations used during augmentation were carefully selected to simulate common types of degradation, including resized crops, colour jitter, and Gaussian blur. These transformations reflect the typical wear and tear of historical manuscripts.

However, during testing, we observed that using all transformations from the self-supervised training stage led to overfitting, reducing classification accuracy. To mitigate this, we opted for a reduced set of augmentations—random resized crop, colour jitter, and Gaussian blur—during the classification phase. This approach balanced enhancing generalisation and preventing the model from becoming overly reliant on specific transformation patterns.

\section{Results and Discussion}

\begin{table*}[htbp]
    \centering
    \caption{Mean classification accuracy for the kNN, SVM, MLP, and prototypical classifiers broken down by samples per class and number of augmentations. Bold values indicate the best-performing non-baseline values achieved in this study. Baseline values, which involved no augmentation, can be seen in the kNN table. The values in the Support/Query samples per class column in the prototypical table represent the data split between the support and query sets required to train this network.}
    \label{tab:classification-results-grid}
    
    \begin{subtable}{0.48\textwidth}
        \centering
        \caption{kNN (including baseline)}
        \label{tab:knn-results}
        \resizebox{\textwidth}{!}{%
        \small
        \begin{tabular}{@{}l|c|*{6}{>{\centering\arraybackslash}p{0.7cm}}@{}}
        \toprule
        \multirow{2}{*}{\makecell[l]{Samples \\ per class}} & \multirow{2}{*}{\makecell{Base-\\line \cite{alfaro-contreras_few-shot_2023}}} & \multicolumn{6}{c}{Number of augmentations} \\
        \cmidrule(l){3-8}
        & & 0 & 1 & 2 & 5 & 10 & 20 \\
        \midrule
        1  & 67.2 & 60.7 & 64.0 & 65.5 & 65.1 & 61.6 & 65.1 \\
        3  & - & 77.1 & 76.8 & 77.9 & 77.4 & 77.0 & 76.3 \\
        5  & 82.0 & 79.8 & 81.4 & 81.6 & 81.4 & 81.9 & 81.9 \\
        10 & 86.9 & 86.1 & 85.9 & 86.5 & 87.5 & 87.7 & 87.3 \\
        \bottomrule
        \end{tabular}%
        }
    \end{subtable}%
    \hfill
    \begin{subtable}{0.48\textwidth}
        \centering
        \caption{SVM}
        \label{tab:svm-results}
        \resizebox{\textwidth}{!}{%
        \large
        \begin{tabular}{@{}l*{6}{>{\centering\arraybackslash}p{1cm}}@{}}
        \toprule
        \multirow{2}{*}{\makecell{Samples per class}} & \multicolumn{6}{c}{Number of augmentations} \\
        \cmidrule(l){2-7}
        & 0 & 1 & 2 & 5 & 10 & 20 \\
        \midrule
        1  & 61.12 & 64.11 & 65.20 & 64.59 & 61.19 & 63.86 \\
        3  & 79.90 & 79.68 & 81.08 & 80.22 & 78.96 & 79.30 \\
        5  & 84.78 & 84.64 & 85.44 & 87.14 & 86.11 & 84.78 \\
        10 & 90.62 & 90.49 & 90.28 & 91.22 & 91.40 & 90.64 \\
        \bottomrule
        \end{tabular}%
        }
    \end{subtable}
    
    \vspace{1em}
    
    \begin{subtable}{0.48\textwidth}
        \centering
        \caption{Multilayer Perceptron}
        \label{tab:rf-results}
        \resizebox{\textwidth}{!}{%
        \large
        \begin{tabular}{@{}l*{6}{>{\centering\arraybackslash}p{1cm}}@{}}
        \toprule
        \multirow{2}{*}{\makecell{Samples per class}} & \multicolumn{6}{c}{Number of augmentations} \\
        \cmidrule(l){2-7}
        & 0 & 1 & 2 & 5 & 10 & 20 \\
        \midrule
        1  & 62.19 & 62.56 & 63.11 & \textbf{64.92} & 64.65 & 64.24 \\
        3  & 81.06 & 81.69 & 81.91 & 81.18 & 81.65 & \textbf{82.54} \\
        5  & 86.79 & 86.67 & 86.48 & 85.92 & \textbf{87.66} & 86.44 \\
        10 & 91.95 & 90.78 & 91.81 & 91.59 & 91.64 & \textbf{92.34} \\
        \bottomrule
        \end{tabular}%
        }
    \end{subtable}%
    \hfill
    \begin{subtable}{0.48\textwidth}
        \centering
        \caption{Prototypical}
        \label{tab:nn-results}
        \resizebox{\textwidth}{!}{%
        \large        
        \begin{tabular}{@{}l*{6}{>{\centering\arraybackslash}p{1cm}}@{}}
        \toprule
        \multirow{2}{*}{\makecell{Support/Query \\ samples per class}} & \multicolumn{6}{c}{Number of augmentations} \\
        \cmidrule(l){2-7}
        & 0 & 1 & 2 & 5 & 10 & 20 \\
        \midrule
        1/2  & 75.96 & 75.44 & 77.40 & 75.06 & 73.43 & 68.68 \\
        2/3  & 84.48 & 84.02 & 82.50 & 77.87 & 80.51 & 81.32 \\
        6/4  & 84.48 & 88.71 & 88.68 & 87.50 & 90.11 & 86.35 \\
        \bottomrule
        \end{tabular}%
        }
    \end{subtable}
\end{table*}

This section presents the classification results, focusing on comparing different classifiers and evaluating the impact of data augmentation. Additionally, the results are contextualised with respect to the kNN-based baseline established by Alfaro-Contreras et al. \cite{alfaro-contreras_few-shot_2023} and provide a discussion on trade-offs between data augmentation and overfitting.

The MLP consistently outperformed the other classifiers across multiple settings, as shown in Table \ref{tab:classification-results-grid}. It achieved an accuracy of 87.66\% with five samples per class, surpassing the kNN baseline by 5.66\% from Alfaro-Contreras et al. (82.0\%) \cite{alfaro-contreras_few-shot_2023}. This demonstrates the strength of MLP’s multi-layered architecture in effectively handling the high-dimensional features extracted by the self-supervised CNN. The MLP's ability to learn complex, non-linear decision boundaries made it particularly well-suited for classifying symbols in this few-shot learning context. This outperformance is consistent with findings from \cite{hospedales_meta-learning_2022} and \cite{snell_prototypical_2017}, where similar architectures were demonstrated to perform well in feature-rich scenarios.

Compared to meta-learning approaches such as Prototypical Networks, the MLP displayed more robust generalization in cases with larger support sets (\(L > 3\)). Prototypical Networks, while effective in low-shot settings (e.g., 75.96\% at \(L = 1\)), showed diminishing accuracy as the augmentation increased, particularly with limited support-query splits, dropping to 68.68\% at \(A = 20\). This suggests that while meta-learning techniques offer advantages in extremely data-scarce contexts, MLP-based methods can scale better as data availability increases. Incorporating additional meta-learning frameworks such as Matching Networks or MAML in future studies could provide further insights into their scalability in OMR tasks.


By contrast, kNN, as highlighted in \cite{alfaro-contreras_few-shot_2023}, performed well in simpler settings but struggled in high-dimensional spaces, where the concept of "nearest neighbors" is weakened by the curse of dimensionality \cite{wang_generalizing_2021}. SVM, although occasionally competitive with MLP, exhibited limitations in handling complex, multi-class scenarios due to its reliance on hyperplanes. These results align with previous work showing that non-linear classifiers, such as MLP, are more adept at capturing intricate patterns in feature-rich datasets.

\begin{figure}[htbp]
    \centering
    \includegraphics[width=1\columnwidth]{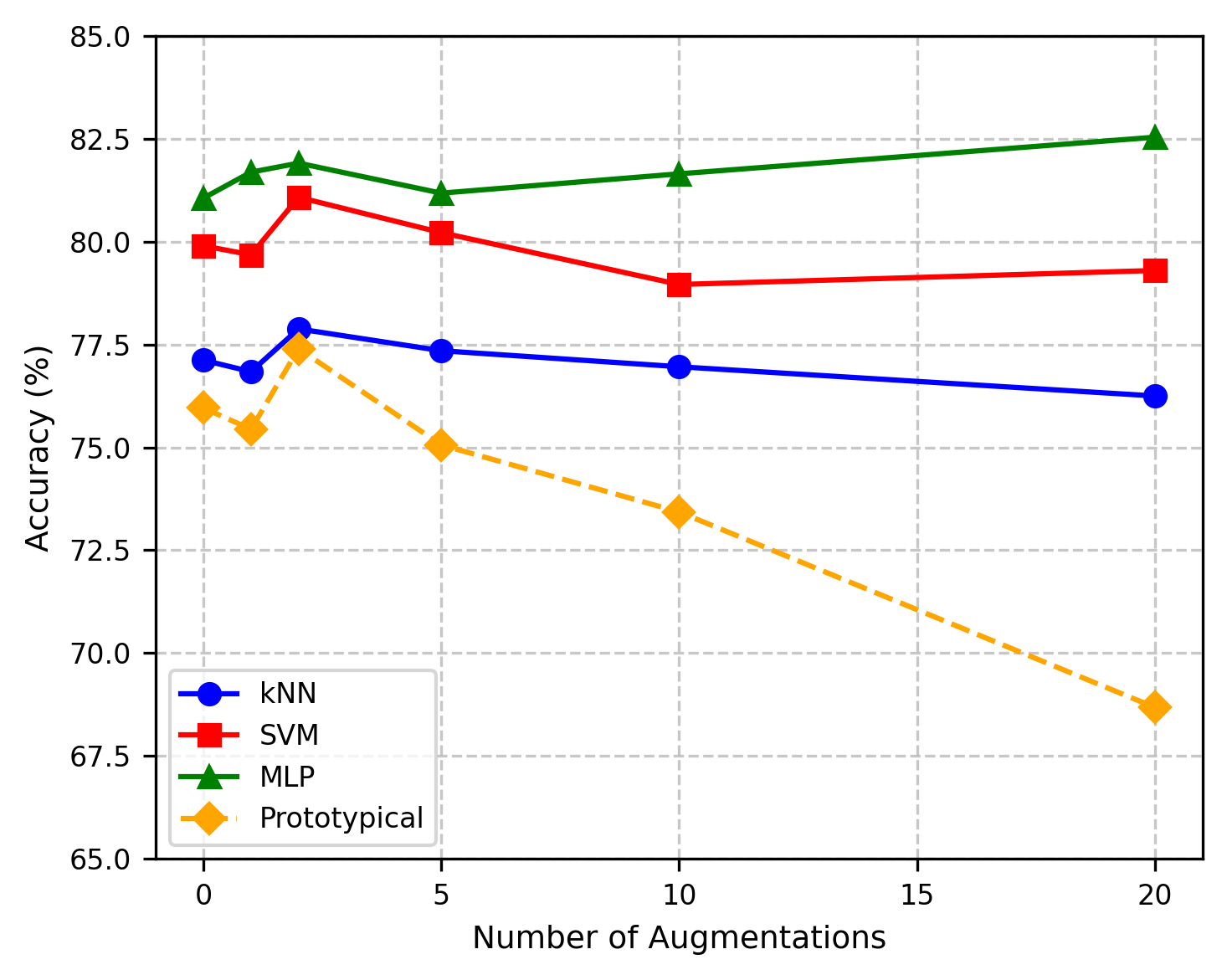}
    \caption{Effect of Augmentations on Classification Accuracy ($K=3$). For the prototypical classifier, the $S=1$ and $Q=2$ configuration is shown.}
    \label{fig:augmentation-effect}
\end{figure}

\begin{figure*}[htb!]
    \centering
    \includegraphics[width=0.76\textwidth]{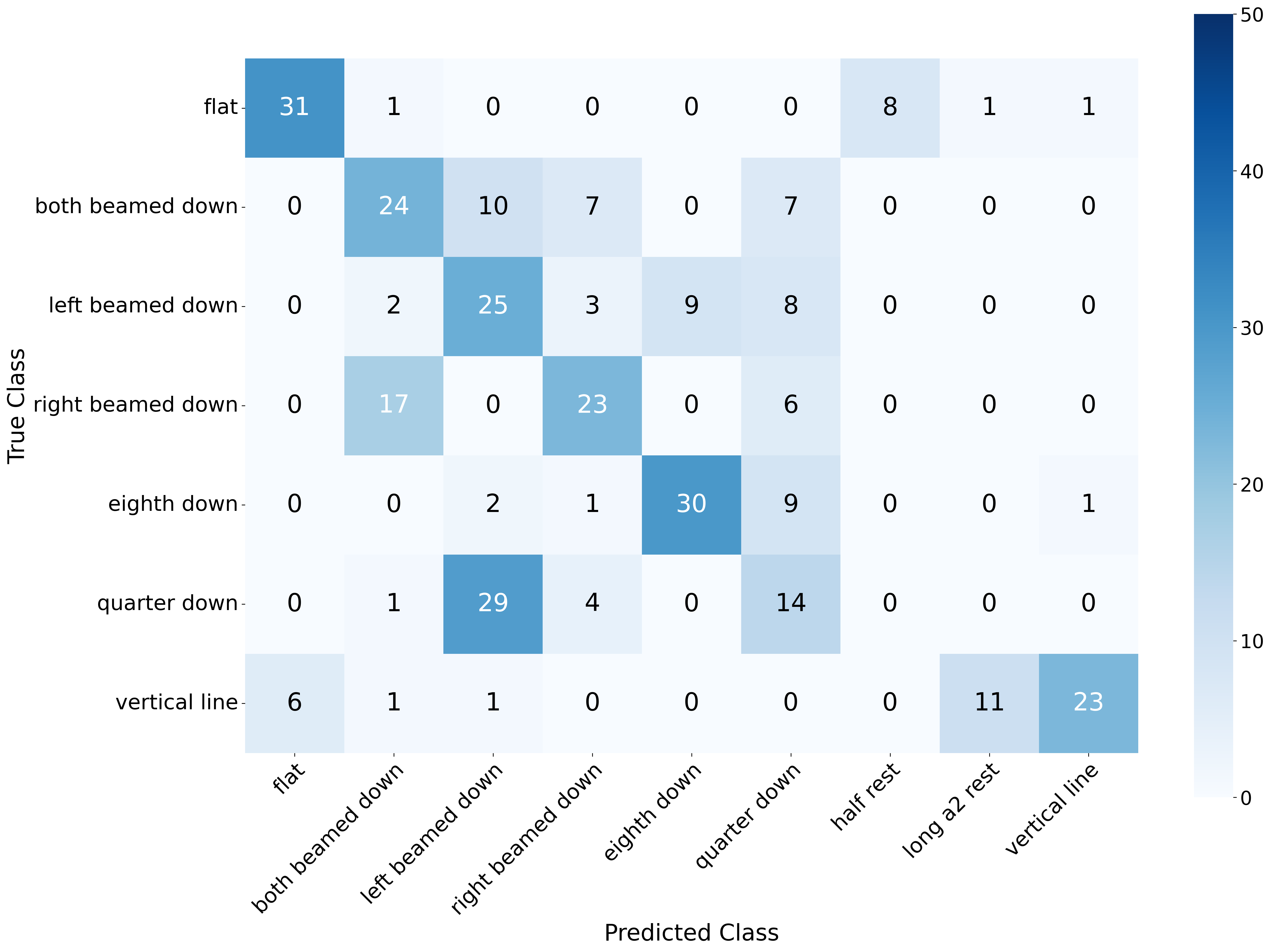}
    \caption{Confusion matrix showing the classification of a subset of classes which were significantly misclassified for all classifiers. This example used the kNN algorithm with $K=10$ and $A=2$. This is an average over five bootstraps.}
    \label{fig:confusion-matrix}
\end{figure*}

\begin{figure}[htbp]
    \centering
    \begin{subfigure}[b]{0.53\textwidth}
        \centering
        \includegraphics[width=\textwidth]{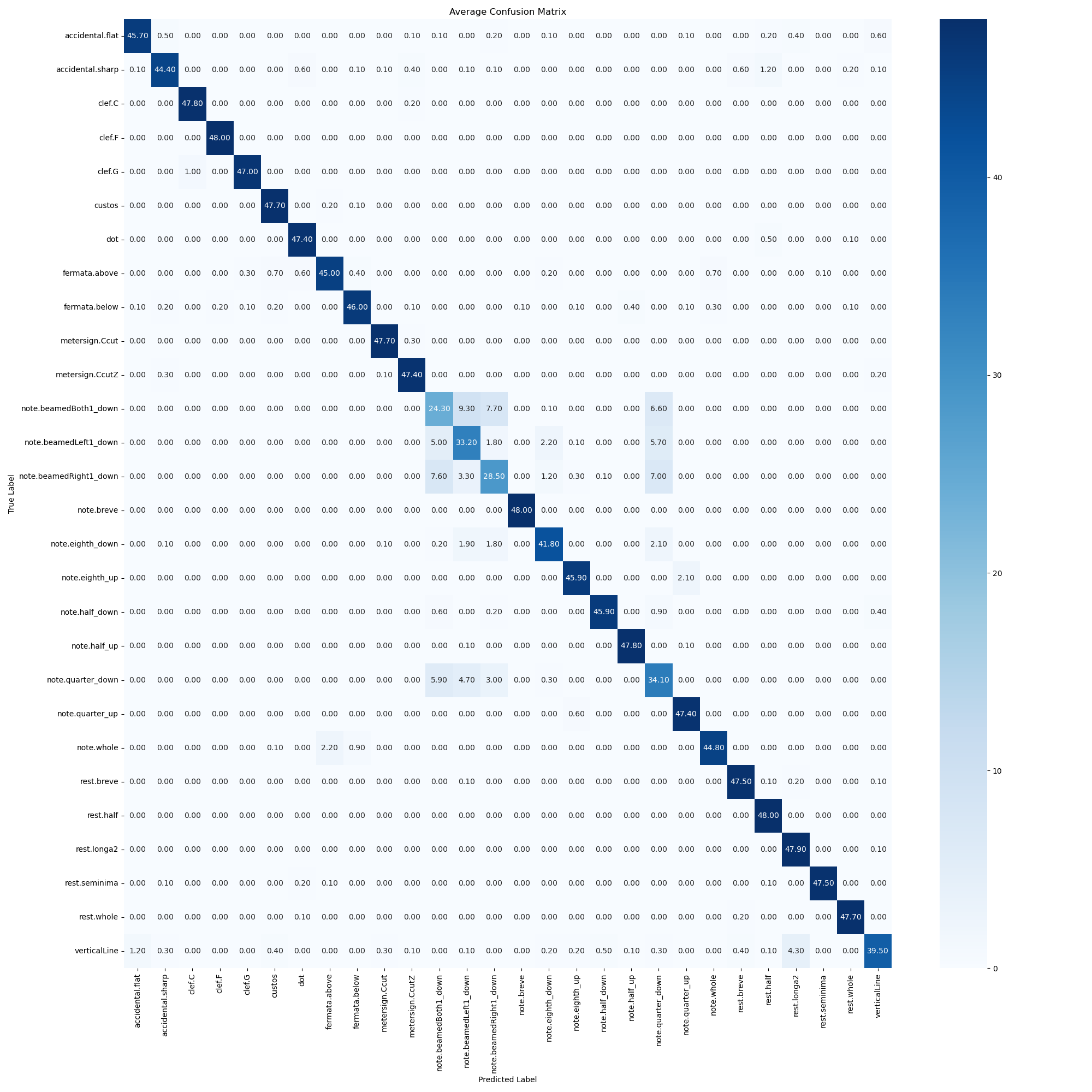}
        \caption{MLP with $K=10$, $A=2$}
        \label{fig:full-MLP-confusion-matrix}
    \end{subfigure}
    
    \begin{subfigure}[b]{0.53\textwidth}
        \centering
        \includegraphics[width=\textwidth]{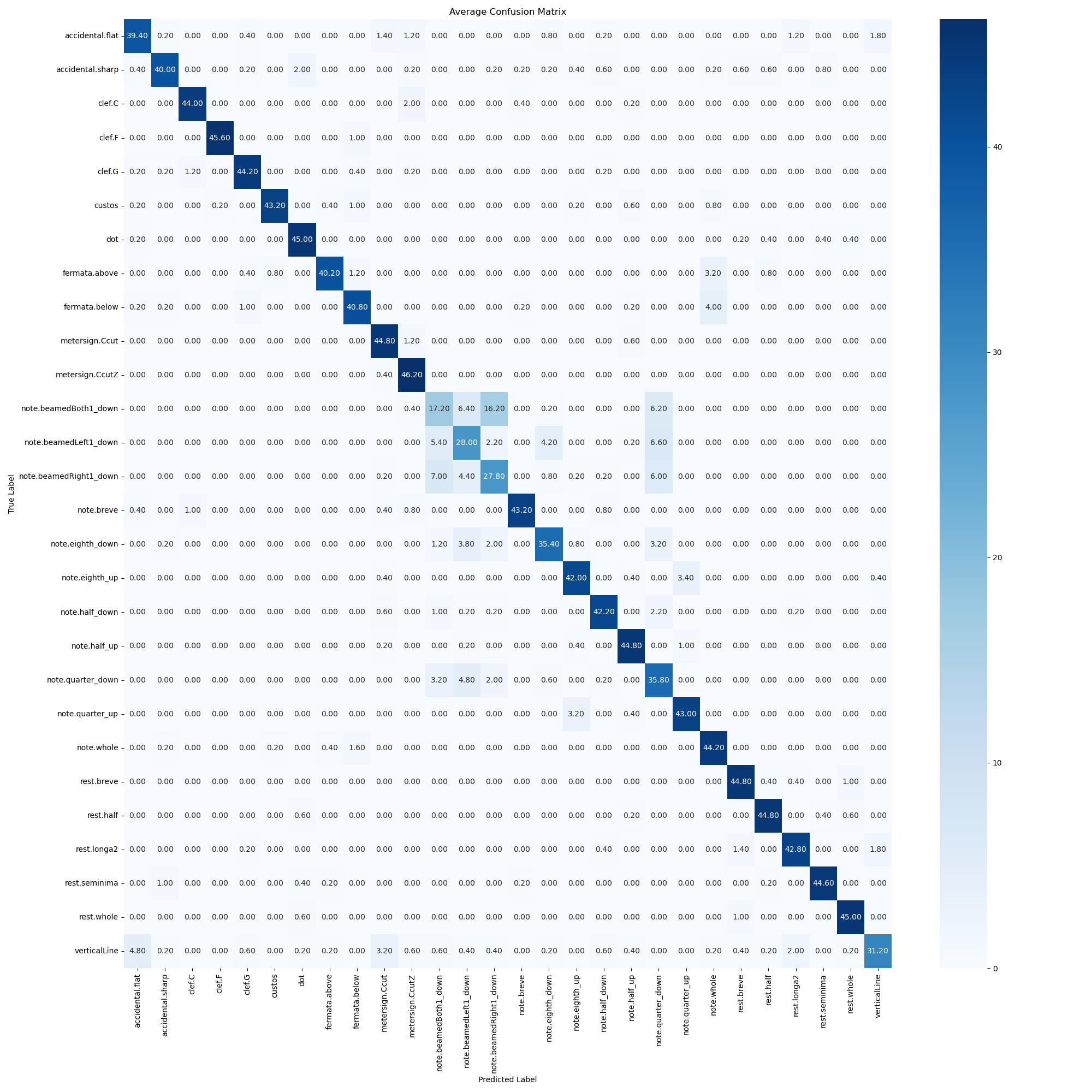}
        \caption{Prototypical classifier with $S=6$, $Q=4$, $A=2$}
        \label{fig:full-proto-confusion-matrix}
    \end{subfigure}
    
    \caption{Confusion matrices for (a) Multilayer Perceptron (MLP) and (b) Prototypical classifier, illustrating the classification performance for various musical symbols in historical manuscripts. The matrices highlight the specific symbols that were frequently misclassified, providing insights into the challenges faced by each model in distinguishing visually similar or degraded symbols.}
    \label{fig:combined-confusion-matrix}
\end{figure}

Data augmentation played a key role in improving the classifiers’ performance, especially for MLP, as shown in Figure \ref{fig:augmentation-effect}. Moderate augmentation levels led to consistent accuracy improvements, but excessive augmentation resulted in diminishing returns for several classifiers, including SVM, kNN, and the prototypical network.

For SVM, over-augmentation may have introduced overly complex decision boundaries, while kNN was likely affected by the cluttered feature space, which hampered the "nearest neighbour" principle. The prototypical network was particularly sensitive to augmentation, with performance dropping from 75.96\% to 68.68\% when too many augmented variations were introduced. This was likely due to the prototypical network's reliance on smaller support sets, making it vulnerable to noise.

However, MLP benefited from higher augmentation levels, showing improvements even with larger sets of augmented samples. This suggests that MLP's deep architecture was better equipped to generalise from augmented data, learning robust features from noisy or altered representations.


Despite the overall performance, all classifiers struggled with certain musical symbol classes, particularly beamed notes, see Figure \ref{fig:confusion-matrix}. These symbols were often confused due to their similar visual structure. This issue indicates a broader challenge in OMR, where subtle visual differences between symbols can be difficult to distinguish with limited training data. These challenges are further compounded in historical sheet music, where inconsistencies in symbol shapes and document degradation create additional obstacles for classifiers. Additional results, including the class-wise confusion matrices for both the MLP and Prototypical classifiers, can be found in the Appendix (refer to Figure \ref{fig:combined-confusion-matrix}).

\section{Conclusion}

This study demonstrated significant improvements in OMR for historical music manuscripts using few-shot learning techniques. The MLP classifier outperformed both the kNN baseline from Alfaro-Contreras et al. and alternative classifiers, showcasing its robustness in feature-rich, low-data scenarios. Additionally, moderate data augmentation enhanced performance, while excessive augmentation highlighted the importance of balancing augmentation strategies to avoid overfitting. The integration of Prototypical Networks provided a comparison with meta-learning approaches, demonstrating complementary strengths but emphasising the scalability of MLP in larger support set scenarios.

Future work will expand evaluations to additional datasets, enabling broader comparisons with other few-shot learning approaches such as MAML and Matching Networks. Furthermore, investigating augmentation strategies tailored to specific symbol complexities, such as adaptive augmentation based on symbol class, could enhance classifier performance in challenging OMR tasks. This continued exploration aims to further bridge the gap between cutting-edge AI techniques and the preservation of cultural heritage through improved OMR systems.

\section*{Acknowledgments}

The authors acknowledge the support of the AI and Music CDT, funded by UKRI and EPSRC under grant agreement no. EP/S022694/1, and our industry partner Steinberg Media Technologies GmbH for their continuous support.





    
    

\vspace{12pt}


\begin{thebibliography}{00}

\bibitem{bainbridge_challenge_2001} 
D. Bainbridge and T. Bell, ``The Challenge of Optical Music Recognition,'' Computers and the Humanities, vol. 35, no. 2, pp. 95--121, May 2001. [Online]. Available: https://doi.org/10.1023/A:1002485918032.
\bibitem{castellanos_preliminary_2023} Castellanos FJ, Gallego AJ, Fujinaga I. A Preliminary Study of Few-shot Learning for Layout Analysis of Music Scores. In5 th International Workshop on Reading Music Systems 2023 Nov (p. 44).
\bibitem{calvo-zaragoza_two_nodate} 
Calvo-Zaragoza J, Rizo D, Quereda JM. Two (Note) Heads Are Better Than One: Pen-Based Multimodal Interaction with Music Scores. InISMIR 2016 Aug 7 (pp. 509-514).
\bibitem{beyer_when_1999} 
Beyer K, Goldstein J, Ramakrishnan R, Shaft U. When is “nearest neighbor” meaningful?. InDatabase Theory—ICDT’99: 7th International Conference Jerusalem, Israel, January 10–12, 1999 Proceedings 7 1999 (pp. 217-235). Springer Berlin Heidelberg.
\bibitem{ding_intelligent_2010}
Ding S, Chen L. Intelligent optimization methods for high-dimensional data classification for support vector machines.
\bibitem{finn_model-agnostic_2017}
Finn C, Abbeel P, Levine S. Model-agnostic meta-learning for fast adaptation of deep networks. International conference on machine learning 2017 Jul 17 (pp. 1126-1135). PMLR.
\bibitem{nichol_first-order_2018}
Nichol A, Achiam J, Schulman J. On first-order meta-learning algorithms. arXiv preprint arXiv:1803.02999. 2018 Mar 8.
\bibitem{hospedales_meta-learning_2022}
Hospedales T, Antoniou A, Micaelli P, Storkey A. Meta-learning in neural networks: A survey. IEEE transactions on pattern analysis and machine intelligence. 2021 May 11;44(9):5149-69.
\bibitem{snell_prototypical_2017}
Snell J, Swersky K, Zemel R. Prototypical networks for few-shot learning. Advances in neural information processing systems. 2017;30.
\bibitem{lim_ssl-protonet_2024}
Lim JY, Lim KM, Lee CP, Tan YX. SSL-ProtoNet: Self-supervised Learning Prototypical Networks for few-shot learning. Expert Systems with Applications. 2024 Mar 15;238:122173.
\bibitem{dhillon_baseline_2020}
Dhillon GS, Chaudhari P, Ravichandran A, Soatto S. A baseline for few-shot image classification. arXiv preprint arXiv:1909.02729. 2019 Sep 6.
\bibitem{shorten_survey_2019}
Shorten C, Khoshgoftaar TM. A survey on image data augmentation for deep learning. Journal of big data. 2019 Dec;6(1):1-48.
\bibitem{chowdhury_few-shot_2021}
Chowdhury A, Jiang M, Chaudhuri S, Jermaine C. Few-shot image classification: Just use a library of pre-trained feature extractors and a simple classifier. InProceedings of the IEEE/CVF International Conference on Computer Vision 2021 (pp. 9445-9454).
\bibitem{souibgui_few-shot_2021}
Souibgui MA, Fornés A, Kessentini Y, Tudor C. A few-shot learning approach for historical ciphered manuscript recognition. In2020 25th International Conference on Pattern Recognition (ICPR) 2021 Jan 10 (pp. 5413-5420). IEEE.
\bibitem{alfaro-contreras_few-shot_2023}
Alfaro-Contreras M, Ríos-Vila A, Valero-Mas JJ, Calvo-Zaragoza J. Few-shot symbol classification via self-supervised learning and nearest neighbor. Pattern Recognition Letters. 2023 Mar 1;167:1-8.
\bibitem{nunez-alcover_glyph_2019}
Nuñez-Alcover A, de León PJ, Calvo-Zaragoza J. Glyph and position classification of music symbols in early music manuscripts. InPattern Recognition and Image Analysis: 9th Iberian Conference, IbPRIA 2019, Madrid, Spain, July 1–4, 2019, Proceedings, Part II 9 2019 (pp. 159-168). Springer International Publishing.
\bibitem{wang_generalizing_2021}
Nuñez-Alcover A, de León PJ, Calvo-Zaragoza J. Glyph and position classification of music symbols in early music manuscripts. InPattern Recognition and Image Analysis: 9th Iberian Conference, IbPRIA 2019, Madrid, Spain, July 1–4, 2019, Proceedings, Part II 9 2019 (pp. 159-168). Springer International Publishing.
\bibitem{song_comprehensive_2022}
Song Y, Wang T, Cai P, Mondal SK, Sahoo JP. A comprehensive survey of few-shot learning: Evolution, applications, challenges, and opportunities. ACM Computing Surveys. 2023 Jul 13;55(13s):1-40.
\bibitem{shatri_optical_2020}
Shatri E, Fazekas G. Optical music recognition: State of the art and major challenges. TENOR Conference 2020, Hamburg, Germany. arXiv preprint arXiv:2006.07885. 2020 Jun 14.
\bibitem{sauvola_adaptive_1997}
Sauvola J, Seppanen T, Haapakoski S, Pietikainen M. Adaptive document binarization. In Proceedings of the fourth international conference on document analysis and recognition 1997 Aug 18 (Vol. 1, pp. 147-152). IEEE.
\bibitem{bardes_vicreg_2022}
Bardes A, Ponce J, LeCun Y. Vicreg: Variance-invariance-covariance regularization for self-supervised learning. arXiv preprint arXiv:2105.04906. 2021 May 11.
\bibitem{shatri2021worms} Shatri E, Fazekas G. DoReMi: First glance at a universal OMR dataset. arXiv preprint arXiv:2107.07786. 2021 Jul 16.
\bibitem{shatri2024kdir} Shatri, E., Fazekas, G. (2024). Knowledge Discovery in Optical Music Recognition: Enhancing Information Retrieval with Instance Segmentation. In \textit{2024 International Conference in Knowledge Discovery and Information Retrieval (KDIR)}.
\bibitem{calvo-zaragoza_handwritten_2019}
Calvo-Zaragoza J, Toselli AH, Vidal E. Handwritten music recognition for mensural notation with convolutional recurrent neural networks. Pattern Recognition Letters. 2019 Dec 1;128:115-21.
\bibitem{rios2024} Ríos-Vila, A., Calvo-Zaragoza, J., Paquet, T. (2024, August). Sheet music transformer: End-to-end optical music recognition beyond monophonic transcription. In International Conference on Document Analysis and Recognition (pp. 20-37). Cham: Springer Nature Switzerland.

\end{thebibliography}
\end{document}